\pdfoutput=1

\documentclass[11pt]{article}

\usepackage{ACL2023}
\usepackage{graphicx} 
\usepackage{times}
\usepackage{latexsym}
\usepackage{amsmath} 
\usepackage{graphicx} 
\usepackage{makecell} 
\usepackage{array} 
\usepackage{multirow}

\usepackage[T1]{fontenc}
\usepackage{booktabs} 
\usepackage[utf8]{inputenc}

\usepackage{microtype}
\usepackage{amssymb}
\usepackage{inconsolata}
\usepackage{graphicx}

\graphicspath{{Images/}}
\title{CoTSRF: Utilize Chain of Thought as Stealthy and Robust Fingerprint of Large Language Models}

\author{
Zhenzhen Ren \quad {\bf GuoBiao Li} \quad {\bf Sheng Li} \\\textbf{}
{\bf Zhenxing Qian} \quad {\bf Xinpeng Zhang} \\
Fudan University \\
\texttt{\{24110240140,20210240200,lisheng,zxqian,zhangxinpeng\}@fudan.edu.cn}
}

\begin{document}
\maketitle
\begin{abstract} 
Despite providing superior performance, open-source large language models (LLMs) are vulnerable to abusive usage. To address this issue, recent works propose LLM fingerprinting methods to identify the specific source LLMs behind suspect applications. However, these methods fail to provide stealthy and robust fingerprint verification.
In this paper, we propose a novel LLM fingerprinting scheme, namely CoTSRF, which utilizes the Chain of Thought (CoT) as the fingerprint of an LLM. 
CoTSRF first collects the responses from the source LLM by querying it with crafted CoT queries. Then, it applies contrastive learning to train a CoT extractor that extracts the CoT feature (i.e., fingerprint) from the responses. Finally, CoTSRF conducts fingerprint verification by comparing the Kullback-Leibler divergence between the CoT features of the source and suspect LLMs against an empirical threshold. Various experiments have been conducted to demonstrate the advantage of our proposed CoTSRF for fingerprinting LLMs, particularly in stealthy and robust fingerprint verification.





\end{abstract}

\section{Introduction}
Recent advanced large language models (LLMs) demonstrate powerful natural language understanding, generation, and reasoning capabilities and have been widely applied in various fields such as 
healthcare  \cite{wang2024pathology}, education \cite{wang2024large}, software development \cite{xia2024agentless}, and scientific research \cite{xia2024agentless}. 
Despite their remarkable success, training a high-performing LLM is not a trivial task, requiring a large scale high-quality data and massive amount of computation resources. 
Fortunately, in actively embracing the ethos of openness, many leading teams in the AI industry have generously released their trained LLMs on open-source platforms such as GitHub and Hugging Face. Notable examples of these open-source LLMs include LLaMA  \cite{touvron2023llama}, Guanaco  \cite{dettmers2024qlora}, and Vicuna  \cite{chiang2023vicuna}, which empowers practitioners with limited resources to conduct further experimentation, fine-tuning, and downstream application development.

For both commercial and ethical reasons, LLM providers typically release their trained LLMs with crafted licenses \cite{cc_noncommercial,gnu_gpl}. These licenses restrict the use of the published LLMs, prohibiting their application in commercial or illegal activities. However, tempted by the enormous profits, some downstream developers may bypass these restrictions, building entities based on open-source LLMs to provide services through APIs, even if these services directly compete with the LLM providers. On the other hand, malicious users may intentionally compromise the LLM's internal alignment mechanisms by fine-tuning, using the LLMs to 
spread harmful content.
Therefore, it becomes a pressing concern for LLM providers to safeguard their released source LLMs against abusive usage (termed as LLM infringement for short) that violates their licenses.

\begin{figure*}[htbp] 
    \centering 
    \includegraphics[width=\textwidth]{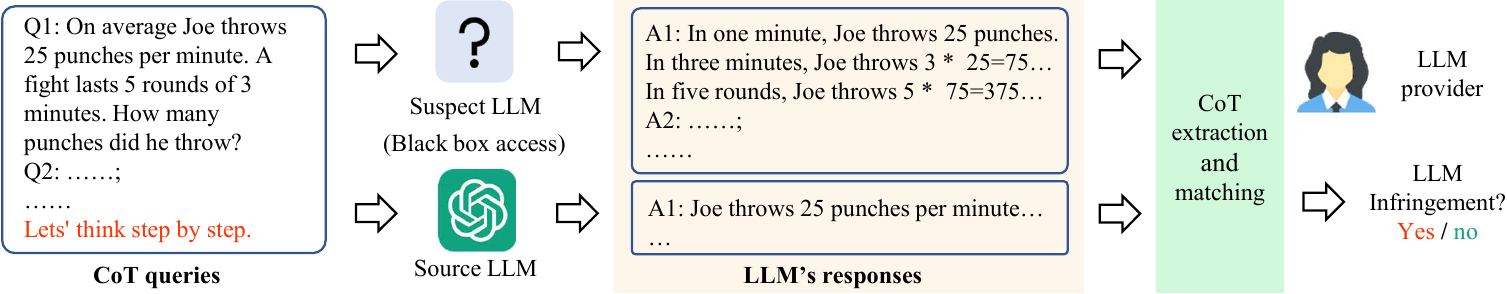} 
    \caption{LLM fingerprint verification process of the proposed method in a black-box access setting, where the model provider only has API access to the suspect LLM.} 
    \label{fig:verification} 
\end{figure*} 

A promising way to address the above issues is model fingerprinting, which non-intrusively extracts the unique features of a model.
Generally, the external manifestation of most of the fingerprinting methods is specific input-output pairs. The inputs are carefully designed to trigger the source model to produce a unique answer while making other benign models to generate different responses. As such, by comparing the unique answer with the response of a suspect API when fed the specific inputs, the model providers could determine whether the model used behind the API is their released one.
Currently, model fingerprinting has achieved significant success in protecting deep neural networks (DNN) \cite{lukas2019deep, zheng2022dnn, peng2022fingerprinting, guan2022you, quan2023fingerprinting}. However, the research on LLM fingerprinting is still in its infancy, possibly due to computational resource limitations and the discrete nature of text data.

A recent proposed LLM fingerprinting method is TRAP  \cite{gubri2024trap}, which combines adversarial prefix \cite{zou2023universal} with queries to induce the source LLM to generate a predefined content. It verifies LLM infringement by comparing the content with the response of the suspect LLM when fed the combined prompts. 
Unfortunately, TRAP is not stealthy, as its added adversarial prefixes is meaningless characters, which disrupt the semantic coherence of prompts. This may alert malicious downstream users and let them to obstruct LLM infringement verification through denial-of-service measures.
On the other hand, TRAP is not robust to output perturbation. If the malicious downstream developers modify the LLM's hyperparameters (e.g., temperature), the infringing LLM (i.e., a copy of the source LLM) may return response that does not match the predefined content.

To bridge this gap, we propose a novel method, namely CoTSRF, which utilizes the \textbf{C}hain \textbf{o}f \textbf{T}hought (CoT) as the LLM's \textbf{S}tealthy and \textbf{R}obust \textbf{F}ingerprint. Our key insight is that the profile of an LLM can be uniquely characterized by its logical reasoning pattern represented by the CoT. Unlike TRAP  \cite{gubri2024trap}, Our CoTSRF verifies the LLM infringement in a stealthy and robust manner, as shown in Fig. \ref{fig:verification}. LLM provider first queries the suspect API using a combination of reasoning questions and a standard CoT prompt (e.g., let's think step by step).
The queried LLM then returns responses that implicitly reveal its logical reasoning patterns. Finally, the provider extracts the CoT features from the responses and compares them with those of the source LLM to identify the LLM infringement. CoTSRF uses the widely adopted CoT prompt to query the infringing LLM, ensuring that it does not alert malicious developers. Moreover, it performs fingerprint matching at the feature level, which is more robust against the output perturbation.

In methodology, CoTSRF begins by obtaining the responses of the source LLM by querying it using reasoning questions and the standard CoT prompt. During this, a High-Temperature Data Augmentation (HTDA) strategy is designed and utilized to generate diverse positive responses that vary in word space but follow the same logical reasoning pattern. Additionally, benign LLMs are used to create negative responses with distinct CoT features. These positive and negative responses are then employed in a contrastive learning framework to train a CoT extractor for accurate CoT feature extraction. In the LLM infringement verification, CoTSRF compares the KL divergence between the CoT features of the source and suspect LLMs against an empirical threshold. 
Various experiments have been conducted to demonstrate the advantages of our proposed CoTSRF for LLM infringement verification. The main contributions are summarized below:

\begin{itemize}
    \item We present the first attempt to leverage CoT as LLM's fingerprint for black-box LLM infringement verification.
    
    \item We propose a novel LLM fingerprinting method CoTSRF that achieves highly competitive results in terms of the stealthy and robustness.
    
    \item We adopt contrastive learning to train a CoT extractor that accurately extracts the LLM's fingerprint from its responses and propose an HTDA strategy to create diverse responses from the LLM.
\end{itemize}

\section{Related Works}
\subsection{Model fingerprinting} 
Model fingerprinting technology non-intrusively extracts the unique features of a source model and uses these features to identify infringing models from benign ones.
It has flourished in protecting DNNs \cite{lukas2019deep, zheng2022dnn, peng2022fingerprinting, guan2022you, quan2023fingerprinting}.
Currently, few attempts have been made in LLM fingerprinting. Zeng \textit{et al.} take the vector direction of the LLM's parameters as the fingerprint and achieve remarkable performance \cite{zeng2023huref}. However, their method requires white-box access to the internal parameters of the suspect LLM during fingerprint verification, making it unsuitable for cases where the malicious developer only provides API access. Gubri \textit{et al.} propose TRAP, which utilizes an adversarial prefix to trigger the source LLM to output a unique answer and takes the mapping of the adversarial prefix and the answer as the fingerprint \cite{gubri2024trap}. Despite supporting fingerprint verification in the black-box setting, TRAP is not stealthy enough to avoid prompt filtering and lacks robustness against output perturbation.

\subsection{LLM Watermarking}
LLM watermarking presents a potential way to address the issue we have highlighted.
It intrusively embeds watermark information within the weights or outputs of the LLM.
Most of the LLM watermarking methods follow a black-box paradigm, where the LLM is fine-tuned to remember distinctive input-output pairs \cite{li2023plmmark, peng2023you, xu2024instructional}. They verify the LLM infringement by comparing the distinctive outputs with the responses of a suspect LLM when fed the specific inputs, which is similar to that of LLM fingerprinting. However, LLM watermarking technology modifies the LLM when embedding the watermark, which inevitably affects the LLM's performance. On the other hand, it cannot adapt to those LLMs that have been released without being watermarked.

\subsection{Chain of Thought} 
Chain of thought (CoT) mirrors LLM's logical reasoning path when solving systematic and complex problems.
Recent researches propose CoT prompting, which  significantly enhances the reasoning abilities of LLMs and makes the output logic of LLMs more reasonable and the results more accurate.
This technology designs CoT prompts to guide the LLM in deconstructing complex problems into orderly sequences of logical steps \cite{wei2022chain}.
Currently, Cot prompting methods could be boradly divided into two campuses: zero-shot CoT and few-shot CoT. The former enables models to generate reasoning steps and solve tasks without any prior examples, effectively leveraging their pretrained knowledge \cite{kojima2022large, wang2023plan, chen2023dynamic, yuan2024instance}. The latter involves generating intermediate reasoning steps for a task, leveraging a handful of examples to enhance model performance \cite{huang2023boosting, song2023llm, liu2024optimizing}.

Recent works have demonstrated that the CoT is LLM's internal attribute, which is highly related to LLM's architecture, training dataset, and training strategy \cite{feng2024towards, liu2024llms}. 
Therefore, in this paper, we argue that CoT could uniquely characterize an LLM and serve as its fingerprint. We then make full use of the popular zero-shot CoT prompting methods to extract LLM's CoT.
Through comprehensive experiments, we empirically demonstrate its effectiveness in identifying specific LLMs and its advantages in stealthy and robust LLM infringement verification.

\section{Problem Formulation}
\subsection{Threat Model}
The threat model of the paper involves two parties: the LLM provider and the malicious downstream developer. The LLM provider releases a source LLM  under a carefully designed license (e.g., a non-commercial license), while the malicious developer ignores these restrictions and builds entities based on the downloaded source LLM to offer profitable services through APIs.

\begin{figure*}[htbp] 
    \centering 
    \includegraphics[width=\textwidth]{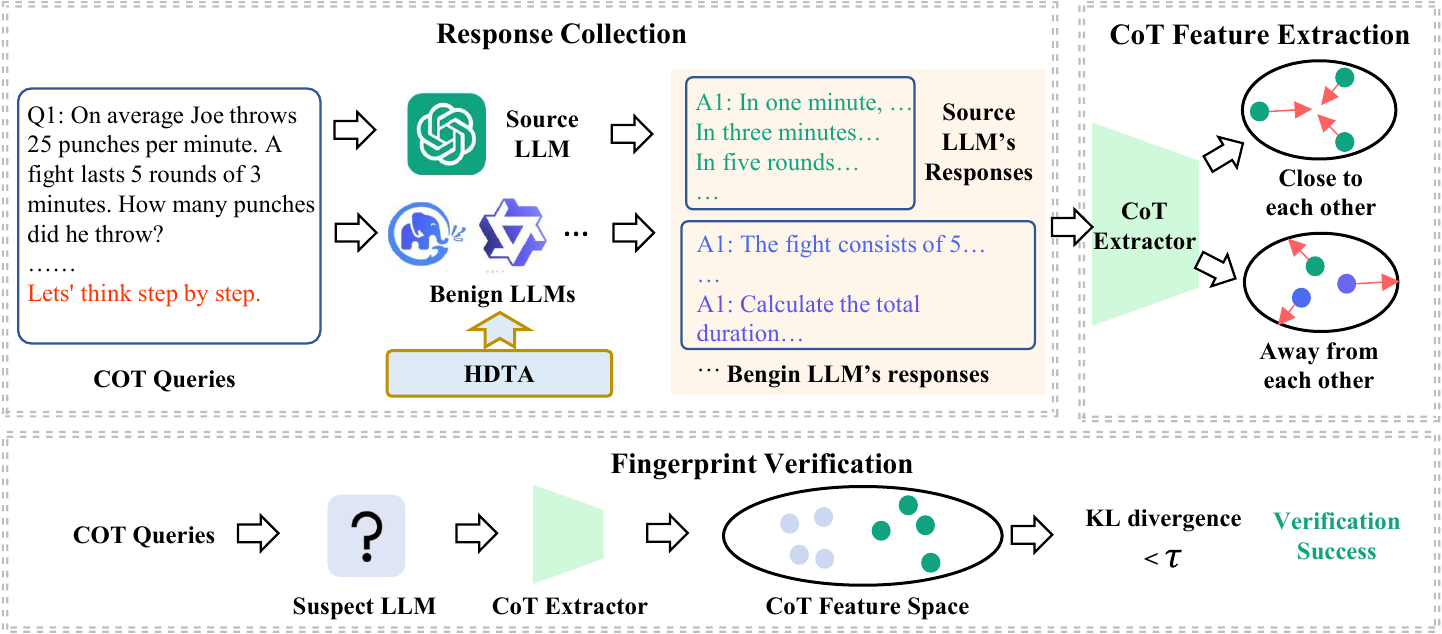} 
    \caption{Framework of the proposed CoTSRF.} 
    \label{fig:main} 
\end{figure*}

The provider’s goal is to identify the infringing LLM that is remotely deployed by the malicious developer. The provider has the following capabilities: 1) white-box access to the source LLM, 2) black-box access to the infringing LLM, and 3) a limited set of queries to conduct fingerprint verification. 
The malicious developer’s goal is to utilize the infringing LLM to provide commercial services without being noticed by the LLM provider.
To prevent the fingerprint verification, the malicious developer conducts the following strategies: 1) at the input level, preset a prompt filter module to check and filter out the odd queries; 2) at the LLM level, fine-tuning the infringing LLM to alter its fingerprint; and 3) at the output level, imposing perturbation to disrupt the fingerprint detection. It should be noted that, the intensity of these strategies must be carefully controlled, as excessive manipulation could compromise the performance of the infringing LLM.

\subsection{Designed Goals}
The design of the LLM fingerprinting method should satisfy the following properties:

\begin{itemize}
  \item [1)] \textbf{Effectiveness}: the fingerprint should accurately identify the infringing LLM;
  \item [2)] \textbf{Reliability}: false positives, where the fingerprint misidentifies a benign LLM released by other providers as an infringing LLM, should be minimized;
  \item [3)] \textbf{Stealthiness}: the fingerprint queries should be normal prompts with coherent semantics to avoid being filtered out;
  \item [4)] \textbf{Robustness}: the fingerprint remains consistent even if the LLM undergoes output perturbation and fine-tuning attacks.
\end{itemize}

\section{Methodology}
    
The flowchart of the proposed method is shown in Fig. \ref{fig:main}. It consists of 
1) a response collection module, which obtains the responses of the LLM by querying it using crafted CoT queries, where a High-Temperature Data Augmentation (HTDA) strategy is designed and used to make LLM generate diverse positive responses that vary in word space but follow the same logical reasoning pattern. It also introduces several benign LLMs to create negative responses with different logical reasoning patterns.
2) a CoT feature extraction module that adopts contrastive learning to train a CoT extractor to extract the logical reasoning path of the responses.
3) An fingerprint verification module, which verifies the LLM infringement by comparing the KL divergence between the CoT features of the source and suspect LLMs against an empirical threshold.
In what follows, we elaborate on each step in detail.

\subsection{Response Collection}

Response collection module involves two types of LLMs: the source LLM \( \mathcal{M}^S \) and a group of benign LLMs \(\{ \mathcal{M}^B_1,  \mathcal{M}^B_2, \cdots, \mathcal{M}^B_K\} \), where $K$ is the total number of benign LLMs.
This module begins by building a set of CoT queries \( Q = \{q_1, q_2, \cdots, q_I\} \), where $I$ is the total number of queries. $q_i \ (1 \leq i\leq I)$ refers to the $i$-th CoT query and is composed of a reasoning question and a standard CoT prompt.
After that, the response collection module collects diverse responses by fedding $Q$ within the source and benign LLMs. 


For \( \mathcal{M}^S \), when fed with \( Q \), it generates \( I \) different responses. Here, we design a High-Temperature Data Augmentation (HTDA) strategy to increase the diversity of the source LLM’s outputs. Specifically, we set the temperature \( T \) of the last softmax layer in \( \mathcal{M}^S \) to a high value and let \( \mathcal{M}^S \) generate \( J > 3 \) different responses for each \( q_i \) in \( Q \). These responses differ in word space but follow the same logical reasoning pattern of \( \mathcal{M}^S \).
After querying \( \mathcal{M}^S \), we obtain a total of \( I \times J \) responses, namely \( R^S = \{ r^s_{1,1}, r^s_{i,j}, \dots, r^s_{I,J} \} \).

By the same token, for the $k$-th benign LLM $\mathcal{M}^B_k$, we obtain a responses set $R^B_k = \{r^b_{1,1,k} \cdots r^b_{i,j,k} \cdots r^b_{I,J,k} \}$, with $r^b_{i,j,k}$ being the $j$-th times response of $ \mathcal{M}^B_k$ for $q_i$.
By querying all the benign LLMs, we obtain $R^B = \{R^B_1, R^B_2, \cdots R^B_K \}$ which consists of $I \times J\times K$ responses.

\subsection{CoT Feature Extraction}
The goal of the CoT feature extraction module is to train a CoT extractor to accurately extract the CoT features from the responses. For reliable fingerprint verification, the extracted CoT features should be similar for the two responses that are both derived from the source LLM but be different when one of them is generated by the benign LLM. To achieve this, contrastive learning with a triplet loss function is adopted to train the CoT extractor. 

For the $i$-th query $q_i$ in $Q$, we consider the two of the responses in $r^s_i$ as positive pairs, and treat the response form  $r^s_i$ and that from $r^b_i$ as negative pairs. Denote the CoT extractor parameterized by $\theta$ as $E_\theta(\cdot)$, we optimize $\theta$ by minimize the following triplet margin loss:

\begin{equation}
\mathcal{L} = \sum_{q_i \in Q} \max\big(0, \|z^a_i - z^p_i\| - \|z^a_i - z^n_i\| + \delta\big),
\end{equation}
where \( z^a_i = E_{\theta}(r^s_{i,j_1})\) is set as anchor CoT feature, and is extracted from the response by the $ \mathcal{M}^S$ for \( q_i \) in $j_1$-th time; 
\( z^p_i = E_{\theta}(r^s_{i,j_2})\) is set as a positive CoT feature, which is extracted from the response by the $ \mathcal{M}^S$ for \( q_i \) in $j_2 (j_2 \neq j_1)$-th time; 
\( z^n_i = E_{\theta}(r^b_{i,j,k})\) is set as a negative CoT feature, which is extracted from one of the responses by the benign LLMs for \( q_i \). \( \| \cdot \| \) denotes the Euclidean distance and \( \delta \) is the margin enforcing a minimum distance between positive and negative pairs.

\subsection{Fingerprint Verification} 



Fingerprint verification module uses CoT queries \( Q = \{q_1, q_2, \cdots, q_I\} \) and source LLM's responses $R^S = \{r^s_{1,1} \cdots r^s_{i,j} \cdots r^s_{I,J} \}$ to verify the LLM infringement. Let's denote the suspect LLM to be verified as $\mathcal{M}^V$. Fed with $Q$, $\mathcal{M}^V$ return a set of outputs $R^V = \{r^v_{1} \cdots r^v_{2} \cdots r^v_{I} \}$, with $r^v_{i}$ denotes the response of $\mathcal{M}^V$ for $q_i$. During the response collection, for $q_i$, $\mathcal{M}^s$ have generated $J$ different responses $\{r^s_{i,1}, r^s_{i,2}, \cdots r^s_{i,J}\}$, from which, we select the first three responses for fingerprint verification. 

Specifically, using trained $E_{\theta}(\cdot)$, we extract the CoT feature vectors of $r^s_{i,1}$ and $r^s_{i,2}$ and measure their distance by  
\begin{equation}
d^{s}_{i} = \|E_{\theta}(r^s_{i,1}) - E_{\theta}(r^s_{i,2})\|.
\end{equation}
We then extract the CoT feature vectors of $r^s_{i,3}$ and $r^v_{i}$ and measure their distance by  
\begin{equation}
d^{v}_{i} = \|E_{\theta}(r^s_{i,3}) - E_{\theta}(r^v_{i})\|.
\end{equation}



Using all the CoT queries \( Q \), we obtain $ D^S = \{d^s_1, d^s_2, \cdots, d^s_I\}$ for source LLM $\mathcal{M}^S$ and $ D^V = \{d^v_1, d^v_2, \cdots, d^v_I\}$ and for verified LLM $\mathcal{M}^V$.
After that, we calculate the distance between the $D^S$ and $D^V$ using the KL divergence with kernel density estimation, as follows:


\begin{equation}
\mathrm{KL}(D^S \| D^V) = \sum_{x \in \mathcal{X}} D^S(x) \log\frac{D^S(x)}{D^V(x)},
\end{equation}
where \( D^S(x) \) and \( D^V(x) \) are probability densities estimated via Gaussian Kernel Density Estimation (KDE) with Silverman's bandwidth rule. The evaluation grid \( \mathcal{X} \) spans \( \big[\min(\mathbf{x}_S, \mathbf{x}_T),\ \max(\mathbf{x}_S, \mathbf{x}_T)\big] \) using 1,000 equally spaced points, where \( \mathbf{x}_S \) and \( \mathbf{x}_T \) denote observed values from each distribution. Numerical stability is ensured by flooring probabilities at \( \epsilon=10^{-10} \).

Finally, we compare \(\mathrm{KL}(D^S \| D^V)\) with an empirical threshold \(\tau\) and identify the verified suspect LLM $\mathcal{M}^V$ as an infringing LLM if \(\mathrm{KL}(D^S \| D^V) \geq \tau\), and as a benign LLM otherwise.

\begin{figure*}[htbp] 
    \centering 
    \includegraphics[width=\textwidth]{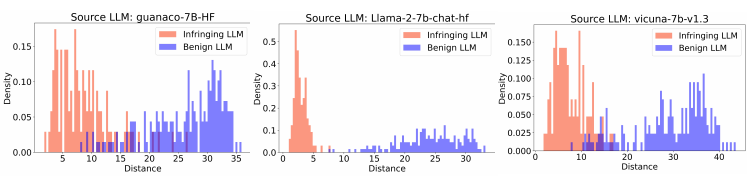} 
    \caption{Distribution of the Euclidean distance between the CoT features of the source LLM and those of the infringing/benign LLM.} 
    \label{fig:distribution} 
\end{figure*}

\begin{table*}[ht]
    \centering
    \caption{Effectiveness and reliability comparison.}
    \label{tab:basic_test}
    \resizebox{.8\textwidth}{!}{%
    \begin{tabular}{l|c|c|c|c|c|c|c|c}
        \hline
        \multirow{2}{*}{Source LLM} & \multirow{2}{*}{$I$} & \multicolumn{5}{c|}{CoTSRF} & \multicolumn{2}{c}{TRAP} \\
        \cline{3-9}
        &  & SI-KL & SB-KL & $\tau$ &TPR & FPR &TPR & FPR \\
        \hline
        \multirow{2}{*}{\texttt{guanaco-7b-HF}} & 50 & 1.5 & 303.6 & \multirow{2}{*}{8.0} & 100.0\% & 0.0\% & - & - \\
        & 100 & 0.8 & 298.6 &&  100.0\% & 0.0\% & 100.0\%   & 0.0\%   \\
        \hline
        \multirow{2}{*}{\texttt{Llama-2-7b-chat-hf}} & 50 & 8.9 & 400.6 & \multirow{2}{*}{85.0} & 99.5\% & 0.0\% & - &-\\
        & 100 & 5.9 & 425.8 & & 100.0\% & 0.0\% & 95.2\%   & 0.2\%   \\
        \hline
        \multirow{2}{*}{\texttt{vicuna-7b-v1.3}} & 50 & 1.7 & 151.2 & \multirow{2}{*}{18.0} &  98.8\% & 0.0\% & - & -\\
        & 100 & 1.1 & 153.5 & & 100.0\% & 0.0\% & 97.0\%   & 0.0\%   \\
        \hline
    \end{tabular}
    }
\end{table*}

\section{Experiments}
\subsection{Experimental Settings}
Three popular open-source LLMs are used in our implementation, including \texttt{Llama-2-7b-chat-hf}, \texttt{vicuna-7b-v1.3}, and \texttt{guanaco-7b-HF}. When one of them is used as the source LLM, the remaining two serve as benign LLMs.
The reasoning questions are derived from the dataset in \citet{wang2023plan}. The standard CoT prompt used to build our CoT queries is: "Let's first understand the problem and devise a plan to solve it. Then, let's carry out the plan and solve the problem step by step." \cite{wang2023plan}.  
The number of CoT queries is either 50 or 100.  
In the HTDA strategy, the temperature \( T \) is set to 1.5, and \( J \) is set to 4.  
The Longformer encoder from \cite{beltagy2020longformer} is adopted as our CoT extractor.  
For each source LLM, we train a unique CoT extractor 3000 epochs using the Adam \cite{diederik2014adam} optimizer.
The hyperparameter \( \delta \) of the Triplet Margin Loss is set to 5.  

For fingerprint verification, the thresholds $\tau$ for \texttt{vicuna-7b-v1.3}, \texttt{Llama-2-7b-chat-hf}, and \texttt{guanaco-7B-HF} are empirically set to 8.0, 85.0, and 18.0, respectively.
Moreover, we introduce \texttt{internlm2\_5-7b-chat} and \texttt{llama3.1-8b-instruct} as unseen benign LLMs to test the reliability of the proposed CoTSRF by verifying its effectiveness in distinguishing those benign LLMs that were not used to train the CoT extractor.
TRAP \cite{gubri2024trap} is used as the benchmark method, which, to the best of our knowledge, is the state-of-the-art LLM fingerprinting method that supports fingerprint verification under a black-box setting. For a fair comparison, we run TRAP with its default settings.

\subsection{Effectiveness}
Effectiveness requires that the fingerprinting method accurately identifies the infringing LLM. Table~\ref{tab:basic_test} presents the verification results of the proposed CoTSRF and the comparison method. In the table, \( I \) represents the number of queries, 
SI-KL/SB-KL represent the KL divergences between the CoT features of the source LLM and the infringing/benign LLM, respectively. 
The True Positive Rate (TPR) measures the accuracy of correctly identifying an infringing LLM, while the False Positive Rate (FPR) indicates the rate of mistakenly identifying a benign LLM as infringing. To obtain the TPR and FPR for Our CoTSRF, we conduct fingerprint verification 100 times for each source LLM.
The TRAP's results are duplicated from its original paper, with “-” indicating no data.

We can see that the difference in values between SI-KL and SB-KL is significant, indicating that the distance between the CoT features of benign LLMs and the source LLM is much greater than that between infringing LLMs and the source LLM. This serves as the foundation of our method’s effectiveness.
For TPR and FPR, our CoTSRF achieves the best results in all cases. Specifically, when $I$ is 100, we provide a 100.00\% TPR across all source LLMs. 
In contrast, TRAP exhibits inferior performance and only provides 95.2\% TPR when taking \texttt{Llama-2-7b-chat-hf} as the source LLM. This highlights the effectiveness of our CoTSRF. 



\begin{table*}[ht]
    \centering
    \caption{Reliability of CoTSRF in identifying unseen benign LLMs.}
    \label{tab:unseen_LLMs}
    \resizebox{.84\textwidth}{!}{%
    \begin{tabular}{l|c|l|c|c|c|c}
        \hline
        Source LLM & $I$ & Unseen benign LLMs & SI-KL & SB-KL & TPR & FPR \\
        \hline
        \multirow{4}{*}{\texttt{guanaco-7b-HF}} & \multirow{2}{*}{50} & \texttt{internLM2.5-7b} & 1.2 & 19.4 & 100.0\% & 3.0\% \\
        &  & \texttt{llama3-8-instruct} & 1.3 & 30.6 & 100.0\% & 0.0\% \\
        & \multirow{2}{*}{100} & \texttt{internLM2.5-7b} & 0.9 & 30.1 & 99.5\% & 0.0\% \\
        &  & \texttt{llama3-8-instruct} & 0.9 & 32.6 & 100.0\% & 0.0\% \\
        \hline
        \multirow{4}{*}{\texttt{Llama-2-7b-chat-hf}} & \multirow{2}{*}{50} & \texttt{internLM2.5-7b} & 8.7 & 318.8 & 100.0\% & 0.0\% \\
        &  & \texttt{llama3-8-instruct} & 8.7 & 333.9 & 100.0\% & 0.0\% \\
        & \multirow{2}{*}{100} & \texttt{internLM2.5-7b} & 5.7 & 329.3 & 100.0\% & 0.0\% \\
        &  & \texttt{llama3-8-instruct} & 5.7 & 331.9 & 100.0\% & 0.0\% \\
        \hline
        \multirow{4}{*}{\texttt{vicuna-7b-v1.3}} & \multirow{2}{*}{50} & \texttt{internLM2.5-7b} & 2.1 & 39.6 & 100.0\% & 0.0\% \\
        &  & \texttt{llama3-8-instruct} & 2.3 & 51.2 & 97.5\% & 0.0\% \\
        & \multirow{2}{*}{100} & \texttt{internLM2.5-7b} & 1.2 & 37.6 & 100.0\% & 0.0\% \\
        &  & \texttt{llama3-8-instruct} & 1.3 & 48.2 & 99.5\% & 0.0\% \\
        
        \hline
    \end{tabular}
    }
\end{table*}

We further visualize the distribution of the Euclidean distance between the CoT features of different types of LLMs using a histogram, as depicted in Fig.~\ref{fig:distribution}. We can observe that, in all cases, the distance between the source LLM and the infringing LLM is significantly lower than that between the source LLM and a benign LLM. This demonstrates that our CoT extractor effectively captures the differences in CoT features between different types of LLMs.


\begin{table}[ht]
    \centering
    \caption{Perplexity of the fingerprint queries of different methods.}
    \label{tab:Stealthiness}
    \resizebox{0.8\linewidth}{!}{%
    \begin{tabular}{c|ccc}
        \hline 
        Methods &  Avg & Min & Max \\ 
        \hline
        Trap    & 16467.4 & 1989.5 &  71995.5  \\
        CoTSRF  & 28.4 & 10.6 & 61.3  \\
        Normal  & 75.7 & 10.2 &  1204.5  \\
        \hline
    \end{tabular}
    }
\end{table}

\subsection{Reliability}
The FPR results in Table~\ref{tab:basic_test} are 0.00\% in all cases, indicating that the proposed CoTSRF can effectively identify the benign LLMs used for training the CoT extractor \( E_\theta{(\cdot)} \). 
To further evaluate the reliability of our method on unseen benign LLMs, we conduct additional tests using \texttt{internlm2\_5-7b-chat} and \texttt{llama3.1-8b-instruct}, which were not included in the training pipeline of \( E_\theta{(\cdot)} \). The results are presented in Table~\ref{tab:unseen_LLMs}.  
We can observe that the SB-KL values remain significantly higher than the SI-KL values, demonstrating a substantial difference between the CoT features of unseen benign LLMs and the source LLM. Moreover, when a large number of CoT queries \( I \) is used, the FPR results remain at 0.00\%.
This indicates that the proposed method can generalize to distinguish unseen benign LLMs from the source LLM, further emphasizing its reliability.

\subsection{Stealthiness}
In this section, we evaluate the stealthiness of TRAP and the proposed CoTSRF. Specifically, we use perplexity \cite{gonen2022demystifying} to measure the semantic coherence of the fingerprint queries generated by different methods. A higher perplexity score indicates lower semantic coherence and a higher probability of being detected and filtered out by a malicious developer.
In our implementation, the standard GPT-2 language model \cite{radford2019language} is used to calculate the perplexity score. The number of queries is set to 100.

Table~\ref{tab:Stealthiness} presents the perplexity scores of the fingerprint queries generated by different methods, with the last row showing the perplexity of normal queries (i.e., reasoning questions without a CoT prompt or adversarial prefix). We can observe that the average perplexity of TRAP’s queries reaches 16467.4, which is significantly greater than that of normal queries (i.e., 75.7.).
In contrast, the perplexity of our CoT-based queries is lower than that of normal queries in terms of the average, maximum, and minimum values. This indicates that the added CoT prompt not only preserves the original logical structure of the queries but also enhances their semantic coherence. These findings demonstrate the stealthiness of our proposed method.

\begin{table*}[ht]
    \centering
    \setlength\tabcolsep{3.5pt}
    \caption{Robustness of CoTSRF against the output perturbation attack in identifying unseen benign LLMs (TPR/FPR).}
    \label{tab:unseen_temp_test}
    \resizebox{\textwidth}{!}{%
    \begin{tabular}{l|ccccccccc}
\hline
    \multirow{2}{*}{Source LLM} & \multicolumn{9}{c}{Temperature $T$ (0.2 - 1.8)} \\
    \cline{2-10}
    & 0.2 & 0.4 & 0.6 & 0.8 & 1.0 & 1.2 & 1.4 & 1.6 & 1.8 \\ 
    \hline
    \texttt{Llama-2-7b-chat-hf}
    & 99.5/0.0 & 99.5/0.0 & 100.0/0.0 & 100.0/0.0
    & 100.0/0.0 & 99.5/0.0 & 100.0/0.0 & 100.0/0.0 & 100.0/0.0 \\
    \texttt{vicuna-7b-v1.3}
    & 95.5/9.5 & 94.0/2.0 & 91.0/5.0 & 96.5/1.0
    & 94.0/2.0 & 97.5/0.0 & 99.5/0.0 & 99.0/0.0 & 98.0/0.0 \\
    \texttt{guanaco-7b-HF}
    & 98.0/0.0 & 100.0/0.0 & 99.0/0.0 & 100.0/0.0
    & 100.0/0.0 & 100.0/0.0 & 100.0/1.0 & 100.0/14.5 & 99.5/85.5 \\
    \hline
    \end{tabular}
    }
\end{table*}

\begin{figure}
    \centering
    \includegraphics[width=0.48\textwidth]{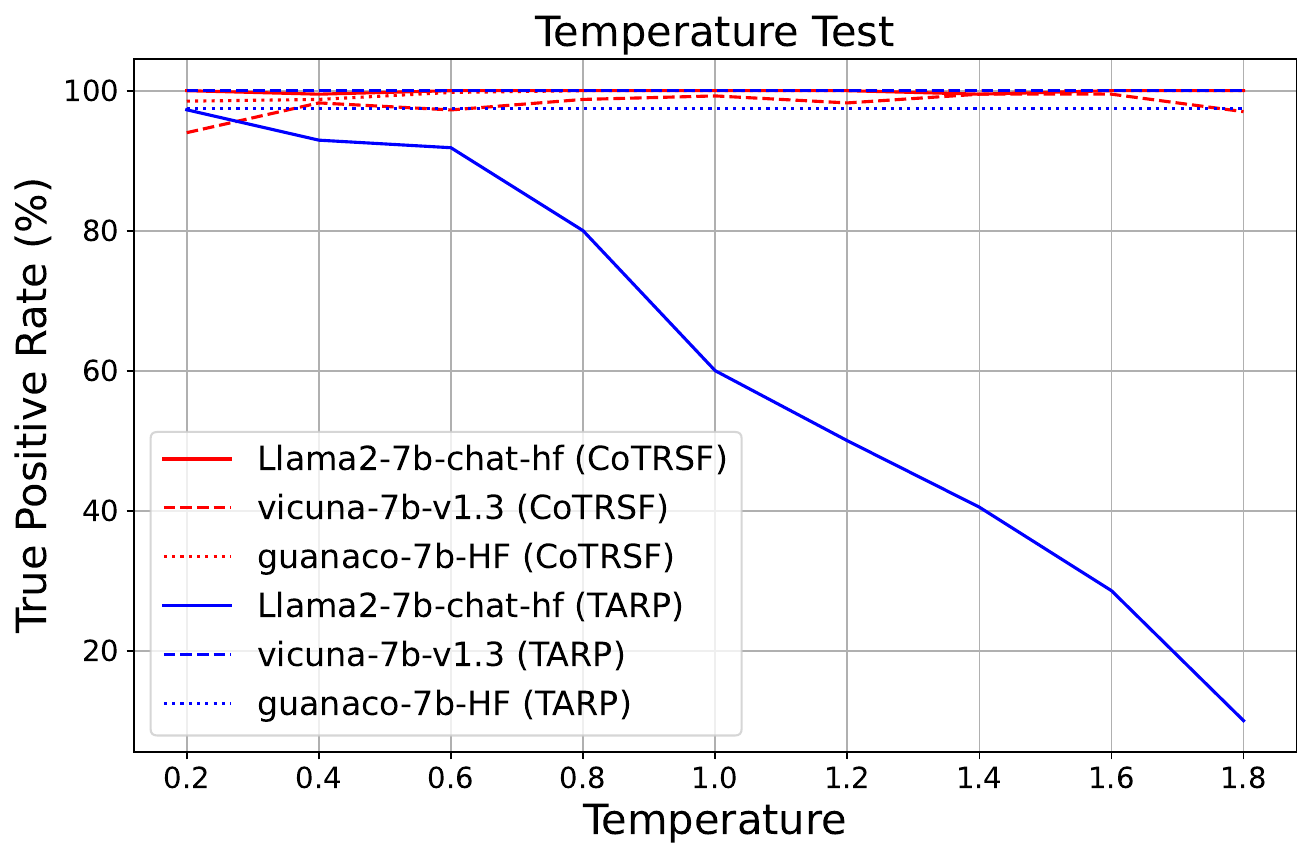} 
  \caption{TPR of CoTSRF and TRAP under different temperature settings (from 0.2 to 1.8).} 
    \label{fig:temperature_base} 
\end{figure}

\begin{table}[ht]
    \centering
    \caption{Robustness against finetuning attacks; (CoT Query number $I=50$).}
    \label{tab:adv_finetune_narrow}
    \resizebox{0.8\linewidth}{!}{%
    \begin{tabular}{c|cc|cc}
        \hline 
        \multirow{2}{*}{Step} &  \multicolumn{2}{c|}{LoRA Rank: 8} & \multicolumn{2}{c}{LORA Rank: 16} \\ 
        \cline{2-5}
        & TPR & FPR & TPR & FPR \\
        \hline
1600  & 100.0\% & 0.0\% &  100.0\% & 0.0\% \\
3200 & 100.0\% & 0.0\% &  100.0\% & 0.0\% \\
4800 & 100.0\% & 0.0\% &  100.0\% & 0.0\% \\
6400 & 100.0\% & 0.0\% &  100.0\% & 0.0\% \\
        
        \hline
    \end{tabular}
    }
\end{table}

In real applications, a malicious developer could implement a perplexity-based filtering module in front of the infringing model to block fingerprint verification by filtering out queries with high perplexity scores. To avoid negatively impacting the performance of the infringing model, they could set the perplexity threshold to the maximum perplexity observed in normal queries (i.e., 1204.5.). Under such conditions, the TRAP method would fail because all of its fingerprint queries would be filtered out. In contrast, all our CoT-based queries can successfully bypass the perplexity-based detection and complete the LLM infringement verification.

\subsection{Robustness}

\subsubsection{Robustness Against Output Perturbation Attack}
After downloading the source LLM, the malicious developer may perturb the LLM's output before returning it to its users to prevent fingerprint verification. Here, we simulate such an attack by adjusting the temperature coefficient \( T \) of the source LLM's last Softmax layer. The values of \( T \) range from 0.2 to 1.8, where a higher \( T \) results in more diverse and random outputs.  

Fig.~\ref{fig:temperature_base} shows the TPR of CoTSRF and TRAP across three LLMs. We can see that CoTSRF consistently achieves TPRs above 94\% across all temperature settings, demonstrating substantial robustness against output perturbation. In contrast, TRAP suffers significant performance degradation at higher temperature settings (i.e., \(T \geq 1.0\)), with TPR dropping below 20\% when \(T\) is 1.8. This highlights CoTSRF’s ability to effectively counter the output perturbation attacks.

We further test CoTSRF’s robustness against output perturbation attack in identifying unseen benign LLMs, including \texttt{internlm2\_5-7b-chat} and \texttt{llama3.1-8b-instruct}, under varying \( T \) and show the results in Table~\ref{tab:unseen_temp_test}. We can see that CoTSRF maintains strong detection performance across moderate temperatures (\( T \in [0.2, 1.4] \)), achieving TPRs above 95\% while keeping FPRs below 2\%. At \( T = 1.8 \), where the output becomes excessively random, CoTSRF's performance begins to degrade for some LLMs (e.g., \texttt{guanaco-7b-HF}). We would like to mention that such extreme temperatures are rare in practical applications, as they may significantly affect the LLM's performance.  

\subsubsection{Robustness against Fine-Tuning Attack}
The malicious developer may also fine-tune the infringing LLM to erase its fingerprint.
To verify the robustness of the proposed CoTSRF against such an attack, we fine-tune \texttt{Llama-2-7b-chat-hf} on the dataset `timdettmers/openassistant-guanaco' \cite{kopf2024openassistant} using Low-Rank Adaptation (LoRA) technology \cite{hu2021lora} on the xTuner framework \cite{contributors2023xtuner}.
The rank of LoRA is set to 8 or 16. The learning rate is set to 1$\times e^{-5}$. Table~\ref{tab:adv_finetune_narrow} gives CoTSRF’s detection performance across different fine-tuning step. As can be seen, even with a high LoRA rank (i.e., 16) and long fine-tuning step (i.e., 4800), our CoTSRF maintains a TPR of 100.00\% and an FPR of 0\%, indicating that CoTSRF is robust to the fine-tuning attack.




\section{Conclusion}

In this paper, we propose a novel LLM fingerprinting method, namely CoTSRF, to identify LLM infringement in a black-box access setting. We take CoT as LLMs' fingerprint, which allows stealthy and robust fingerprint verification. Specifically, we first collect the responses of the source LLM by querying it using crafted CoT queries. During this, a High-Temperature Data Augmentation (HTDA) strategy is proposed to boost the diversity of the responses. We then employ a contrastive learning framework with triplet margin loss to train a CoT extractor for accurate CoT extraction. Various experiments demonstrate the advantages of our method for verifying LLM infringement, achieving satisfactory performance in terms of effectiveness, reliability, stealthiness, and robustness.

\section{Limitation}

While our proposed CoTSRF method demonstrates strong performance in fingerprinting and identifying a wide range of LLMs, it is not without limitations. These limitations primarily stem from highly contrived corner cases and the need for broader validation across diverse model scales and architectures. Below, we discuss these challenges in detail.  

\textbf{1. Challenges in Highly Contrived Corner Cases}
The method faces difficulties in scenarios where two LLMs exhibit extreme architectural and training homogeneity. For instance, if two entities independently train models using identical open-source architectures (e.g., LlaMA) with precisely replicated training protocols—same data sources, identical preprocessing—the resulting models' reasoning pathways could become nearly indistinguishable. This would reduce the effectiveness of CoT-based fingerprinting, leading to a decline in the reliability of our method, as it would struggle to differentiate between the source model and the benign model. However, such scenarios are exceptionally rare in practice due to the low probability of two entities independently replicating the exact same training pipeline. 

\textbf{2. Need for Broader Validation}
While our experiments demonstrate promising results on several models under 10B parameters, the generalizability of CoTSRF requires further validation across a broader range of LLM architectures and scales. The rapid evolution of LLM architectures necessitates testing on larger and more diverse model families. Future work should extend evaluations to larger-scale models (e.g., 70B+ parameters) and emerging paradigms such as mixture-of-experts and multimodal architectures.

\section*{Acknowledgements}

\bibliography{main}
\bibliographystyle{acl_natbib}

\appendix

\section{Example Appendix}
\label{sec:appendix}

This is a section in the appendix.

\end{document}